\documentclass[12pt]{iopart}
\usepackage{epsfig}
\usepackage{graphics}
\usepackage{amssymb}
\usepackage{cite}
\usepackage{caption}
\usepackage{subcaption}

\def\eq#1{(\ref{#1})}

\begin{document}
\title[]{Holographic complexity and the Hubble tension: a quantum gravity portrayal for the large scale structure of the cosmos.}

	\author{Carlos Silva \footnote{Present address:
	Instituto Federal de Educação Ciência e Tecnologia do Ceará (IFCE),  Campus Tianguá,
	Av Tabelião Luiz Nogueira, s/n - Santo Antônio, Tianguá, Ceará, Brazil.}}

\ead{carlosalex.phys@gmail.com}

\begin{abstract}

In this paper, we propose a relationship between the so-called Hubble-Lema\^{i}tre constant $H_{0}$ and the holographic complexity related to the emergence of spacetime in quantum gravity. Such a result can represent an important step to understanding the Hubble tension by introducing a quantum gravity perspective for cosmological observations, regarding the degree of quantum complexity we measure around us.

\end{abstract}


\maketitle

\section{Introduction}

In recent years, a controversy has arisen in cosmology, known as the Hubble tension \cite{Verde:2019ivm, DiValentino:2021izs, Freedman:2021ahq}.
This problem has its origin in a non-concordancy between early-universe measurements of the so-called Hubble-Lema\^{i}tre constant $H_{0}$ provided by the Planck collaboration \cite{Planck:2018vyg}, and late-time observations provided by SHOES \cite{Riess:2016jrr}.
Such a discrepancy has been confirmed by several other late-time data \cite{Riess:2016jrr, Wong:2019kwg, Freedman:2019jwv, Yuan:2019npk, LIGOScientific:2019zcs, Pesce:2020xfe, Soltis:2020gpl, Riess:2020fzl}, which have shown a clear distinction from the value provided by the Planck satellite.

The early-universe result is based on the analysis of the Cosmic Microwave Background radiation (CMB), which was formed when the universe was in its childhood. On the other hand, late-time  measurements are based on several observations related to objects formed in epochs when the universe was older and more complex, such as Cepheids, quasars, and red giant stars.

Putting it in numbers, the early universe result has provided a value of $H_{0} = 67.4 \pm 0.5 \; km \; s^{-1}Mpc^{-1}$. On the other hand, the late-time universe measurements have provided values for $H_{0} \geq 70$. For example, the latest SHOES measurement gives $H_{0} = 73.2 \pm 1.3 \; km \; s^{-1}Mpc^{-1}$ \cite{Riess:2020fzl}. Moreover, several middle-time observations have provided intermediate values \cite{eBOSS:2020yzd, eBOSS:2020tmo}.

In statistical terms, the difference between early and late-time results amounts to $4\sigma$ to $6\sigma$ depending on the datasets considered \cite{DiValentino:2021izs}. Such a discrepancy has cast doubt on the very constancy of $H_{0}$, and is considered sufficient to establish a crisis in cosmology, asking for new physics \cite{Vagnozzi:2019ezj, Krishnan:2020vaf, Colgain:2022rxy, Vagnozzi:2023nrq, Montani:2023xpd, Montani:2023ywn, Schiavone:2022wvq}.

Several proposals have been suggested to mend the $\Lambda$CDM model and explain the different values observed of $H_{0}$.
An extensive list of such proposals can be found in \cite{DiValentino:2021izs}.
In front of these efforts, it has been argued that the Hubble tension must, in fact, consist of a crisis just established in the standard model of cosmology. In this case, a deeper breakdown in our understanding of the cosmos becomes necessary to solve it.

A forthcoming quantum version of cosmology, based on a still unknown theory of quantum gravity, can promote such a breakdown by reformulating our view about one of the most fundamental aspects of physical reality: spacetime itself.
However, so far, no satisfactory proposal to solve the Hubble tension has been introduced by any candidate theory of quantum gravity.

In the present paper, we propose a possible explanation for the origin of the Hubble tension by considering new results on the nature of spacetime in quantum gravity introduced in \cite{Silva:2020bnn, Silva:2023ieb}. According to such results, spacetime must not be fundamental but emergent from quantum correlations without correlata, a reminiscent idea from Mermin's interpretation of quantum mechanics \cite{Mermin:1996mr, Mermin:1998cg}. In such a context, we shall find that $H_{0}$ must be related to which must arise as an important element to understand the quantum aspect of the cosmos: the quantum complexity associated with the emergence of spacetime.

Consequently, the discrepancies between the $H_{0}$ values, when one considers the early and late-time measurements, can be attributed to a natural feature of the cosmos: the evolution of its complexity throughout time. Such a result may not only shed light on the Hubble tension but also introduce a new perspective for cosmological observations: in terms of holographic quantum complexity, we measure around us.

The paper is organized as follows: in sections \eq{sec2} and \eq{sec3}, we review the main aspects of the theory introduced in \cite{Silva:2020bnn, Silva:2023ieb}. In section \eq{sec4}, we address some basic concepts related to holographic quantum complexity. In section \eq{sec5}, we demonstrate that $H_{0}$ can be related to holographic complexity required for the emergence of spacetime. In section \eq{sec6}, we show that the relationship between $H_{0}$ and holographic complexity can reproduce the results given for $H_{0}$ by early, middle, and late-time measurements. Section \eq{conc} is devoted to conclusions and discussions.

\section{The holographic principle as a prelude to the emergence of spacetime} \label{sec2}

A forthcoming quantum theory of gravity must provide a radical change in our comprehension of the world. However, possibly the most disruptive lesson quantum gravity must bring us is that related to the role of spacetime itself in our description of physical reality.

In this sense, the holographic principle has been pointed out as a possible guide to quantum gravity \cite{tHooft:1993dmi, Susskind:1994vu, Bousso:2002ju}. According to such a principle, the gravitational dynamics in a region of spacetime can be completely described by
a quantum theory living on its boundary.

The most successful form of the holographic principle is the so-called AdS/CFT correspondence consisting of a non-perturbative approach to superstring theory \cite{Maldacena:1997re}. In its original formulation, the AdS/CFT correspondence establishes a duality between a type IIB string theory living in a $AdS_{5} \times S^{5}$ bulk and a conformal field theory (CFT) living on the bulk boundary.

From the necessity of aligning the AdS/CFT ideas with the gravitational dynamics present in our four-dimensional universe, it has been considered its complementarity with the Randall-Sundrum II (RSII) scenario \cite{Randall:1999vf, Gubser:1999vj, Verlinde:1999fy, Duff:2000mt}, another important insight in the string theoretical setup. By taking this way, our four-dimensional universe, viewed in the AdS/CFT framework as localized on the AdS boundary, turns out to be described by a stack of coincident Randall-Sundrum II branes, and the CFTs must be coupled to gravity.

Such a complementarity between the AdS/CFT  and the Randall-Sundrum frameworks has given birth to a new version of the AdS/CFT correspondence, which has been proposed
by Emparan, Kaloper, and Fabri \cite{Emparan:2002px}, and independently by Tanaka \cite{Tanaka:2002rb} in 2002.
It was named by Gregory as the Black Hole Holographic Conjecture (BHHC) \cite{Gregory:2004vt, Gregory:2008rf}, and
proposes that the gravitational theory living in the AdS bulk, described by a type IIB string theory, must be holographically related not only to a CFT but to a semiclassical gravitational theory (gravity + quantum corrections) living on the AdS boundary. In such a context,  bulk higher-dimensional effects must be translated as quantum corrections to the boundary gravitational theory.

The BHHC has opened up the possibility of interesting investigations on the nature of the gravitational degrees of freedom belonging to the boundary theory in the AdS/CFT framework.
For example, by applying its ideas to a cosmological scenario, it has been demonstrated that quantum-corrected Friedmann equations similar to  Loop Quantum Cosmology (LQC) ones can be induced on the AdS boundary \cite{Silva:2020bnn}.

Since quantum corrections arising in such a context must correspond to the holographic dual of higher-dimensional effects in the AdS space, such a result has raised the interesting possibility of connecting superstring theory living in the AdS bulk to a quantum gravitational theory living on the boundary.

In fact, it has been demonstrated that LQG polymer quantization can be applied to the boundary gravitational theory in such a context. In this case, the quantum degrees of freedom belonging to this theory must correspond to $U(1)$ quantum holonomies giving rise to quantum geometrical states $\psi(x) = e^{ipx}$ defined on the AdS boundary in the following way:

\vspace{5mm}

\begin{equation}
\hat{h}_{\pm}\psi (x) = e^{ip\sqrt{\Delta}}\psi (x) = e^{ip(x+ \sqrt{\Delta})} = \psi(x+ \sqrt{\Delta}). \label{holonomy-action}
\end{equation}

\vspace{5mm}

Such results introduce a new scenario where braneworld geometrical degrees of freedom can be associated with a polymer structure, defined by a graph in the form of a regular lattice:

\begin{equation}
\gamma_{\sqrt{\Delta}} = \{ x \in \mathbb{R} \mid x = n\sqrt{\Delta}, \forall n \in \mathbb{Z} \}\;. \label{regular-latt}
\end{equation}

\vspace{5mm}

\noindent where $\sqrt{\Delta}$ consists of the lattice spacing parameter.

It establishes a reinterpretation of branes as  $U(1)$ quantum polymer structures, and a possible connection between the main approaches to quantum gravity nowadays can be traced by connecting fundamental objects in string theory and LQG: branes and polymer structures.

On the other hand, we note that even though the holonomies above are similar to that we have in LQG to describe a quantum spacetime, in the scenario introduced in \cite{Silva:2020bnn}, the quantization parameter defining such objects is related to the brane tension and, consequently, to the closed string coupling \cite{Becker:2007zj, Zwiebach:2004tj}.
In fact, the parameter $\Delta$ in Eq. \eq{holonomy-action} correspond to a quantum of area parameter given by \cite{Silva:2020bnn}:

\begin{equation}
\Delta =  \frac{12\pi}{\sigma} =  96\pi^{4}\alpha'^{2}g_{s}  = \frac{24 \pi^{3}L^{4}}{N} \;, \label{delta-g1}
\end{equation}

\noindent where $\sigma$ is the brane tension,  $\alpha'$ is the Regge slope, $g_{s}$ is the closed string coupling, $L$ is the AdS curvature radius,  and  $N$ corresponds to the number of branes sourcing gravitationally the AdS bulk geometry \cite{Becker:2007zj, Zwiebach:2004tj}.
We note, from Eq. \eq{delta-g1}, that the quantum of area $\Delta$ depends on the energy scale (or equivalently the length scale) used to probe spacetime geometry.


\vspace{5mm}

The results above have been useful in addressing interesting problems as those related to the big bang singularity in $AdS/CFT$ correspondence \cite{Bak:2006nh, Engelhardt:2015gla}, and the Immirzi ambiguity in LQG \cite{Rovelli:1997na}. In this context, it has been also demonstrated that a positive cosmological constant can be induced on the AdS boundary, which has open up the possibility of explore possible observational evidences of the AdS/CFT correspondence.

Amidst all these results, the main lesson we have is that the way branes must be combined to source gravitationally the AdS geometry will depend on the rules governing quantum polymer structures, similar to those that appear in LQG. However, at this point, some care is needed in interpreting the use of LQG techniques in the context of the AdS/CFT correspondence. In this sense, the existing LQG literature offers two possibilities.

The first one is that the use of LQG techniques will provide the quantization of the AdS boundary geometry. However, it has been demonstrated that the LQG quantization techniques not only quantize spacetime but also modify it \cite{Bojowald:2011hd}. In this case, one should not have an AdS spacetime anymore after the application of the LQG quantization methods to the AdS boundary, and the bridge between LQG and string theory would be broken.

The second, and most radical possibility offered by the LQG literature, is that the LQG quantization procedure would not only quantize spacetime, but it would also dissolve it into a pre-geometric structure that lives beyond spacetime itself, and from which spacetime ought to emerge.

Such a perspective is in tune with some AdS/CFT ideas that propose spacetime must be an emergent aspect of physical reality \cite{Ryu:2006bv, Ryu:2006ef, Nishioka:2009un, Takayanagi:2012kg, VanRaamsdonk:2009ar, VanRaamsdonk:2010pw}.
In fact, the AdS/CFT framework provides the most well-understood example for spacetime emergence in quantum gravity by proposing that the AdS bulk must be emergent from quantum entanglement related to quantum degrees of freedom living on the AdS boundary.

However, the fundamental aspects of how bulk spacetime and its gravitational dynamics emerge from the dual theory remain shrouded in many mysteries. For example, it has been suggested that entanglement is not enough to explain the emergence of spacetime, but quantum complexity must also play a fundamental role in this process. As we shall see, the theory introduced in \cite{Silva:2020bnn} may bring us new insights into this issue.

\section{A holographic prescription to the emergence of spacetime in quantum gravity.} \label{sec3}

In order to build a background-independent approach to quantum gravity, it has been proposed that spacetime must emerge from quantum information codified in a set of quantum reference frames connected by quantum correlations \cite{Girelli:2005ii, Castro-Ruiz:2019nnl, Smith:2020zms}.

The notion of a quantum reference frame arises in a quantum mechanical description of the world as related to the universality of the theory: both observers and observed must satisfy the same natural laws. In this case, if the quantum description of the world is universal, one would expect that reference frames (observers) should obey the laws of quantum mechanics, corresponding in this way to quantum systems, such as quantum particles.

However, what kind of system could be considered as a quantum reference frame in the context of quantum gravity? Taking into account the string theoretical setup, a brane is a physical object that generalizes the notion of a point particle. In this way, one can naturally suggest that these objects can assume the role of quantum reference frames in a quantum gravitational context.

By taking such a concern, the results found in \cite{Silva:2020bnn} can provide the missing ingredient to conceive branes as quantum reference frames: that these objects can be seen as $U(1)$ quantum holonomy structures. In this case, one can see branes as quantum clocks, i.e., temporal quantum reference frames to the emergence of spacetime in quantum gravity \cite{Girelli:2005ii, Castro-Ruiz:2019nnl, Smith:2020zms}.

However, before proceeding, we note that a peculiar detail related to the scenario introduced in \cite{Silva:2020bnn} arises in this discussion: the holonomic structure of branes can be used not only to establish these objects as quantum reference frames but also to weave the entanglement network between them. Such a conclusion is rooted in the fact that quantum holonomies $\hat{h}_{ij}$ consist of unitary transformations connecting two quantum geometrical states, belonging (in general) to two different Hilbert spaces, through entanglement \cite{Baytas:2018wjd, Bianchi:2018fmq, Mielczarek:2018jsh}:

\begin{equation}
\hat{h}_{ij} \; : \; \mathcal{H}_{i} \rightarrow \mathcal{H}_{j}\;, \label{correlation1}
\end{equation}

\noindent where

\begin{equation}
\hat{h}_{ij}\hat{h}_{ij}^{\dagger}  = \mathbb{I} \;. \label{correlation2}
\end{equation}

\vspace{1cm}

\noindent In this context, if we label the $\mathcal{H}_{i}$s as the Hilbert spaces of $N$ branes, in the equations above, the indices $i,j$ will range as $i,j = 1,..., N$, and the holonomies $\hat{h}_{ij}$ will be promoted to $U(N)$ unitary matrices.

However, the important detail is that there is no fundamental distinction between the holonomies describing branes as quantum reference frames and the holonomies establishing the quantum correlations between them. Note that, in such a construction, one could arbitrarily label some holonomies to correspond to those frozen to define the brane quantum reference frames, while taking the others to correspond to the correlations between such objects.
Consequently, even what has been called a quantum reference frame can be understood fundamentally as a quantum correlation, and vice versa.

In this case, as it was pointed out in \cite{Silva:2023ieb}, it is possible to take the universality of the theory one step further by introducing a scenario where there exists a unique kind of quantum object: quantum correlations. In this way, it is possible to introduce a Structural realistic interpretation to quantum gravity \cite{Ladyman:1998}, where only quantum correlations are considered as fundamental, and quantum refence frames, as well as spacetime itself, would be seen as completely emergent from the combinatorial rules governing the structure of such correlations. Such ideas are reminiscent of the Mermin's interpretation of quantum mechanics, where only quantum correlations must correspond to the elements of physical reality \cite{Mermin:1996mr, Mermin:1998cg}.

Such a perspective, introduced in \cite{Silva:2023ieb}, can be useful to understand some of the most fundamental aspects related to the emergence of spacetime in quantum gravity. For example, according to the usual form of the holographic principle, spacetime geometry must emerge holographically from a quantum theory living in one spatial dimension lower. It corresponds to the usual AdS/CFT framework for the emergence of spacetime.

However, a conceptual problem has been pointed out in such a construction: in the situation where the bulk emerges from the boundary, the latter should be more fundamental than the former. It is difficult to reconcile with the fact that, in the AdS/CFT scenario, the duality between the boundary and the bulk theories is exact and thus symmetrical \cite{Teh:2013xka, Dieks:2015jja, DeHaro:2016lzy, Bain:2020ajf}.

This kind of conceptual barrier is rooted in the fact that it is a challenge to completely leave the idea of the fundamental existence of spacetime in physics. In fact, how could physics exist beyond spacetime, and how could things exist and become entangled, without some loci where and when they happen and change? In this case, in the usual form of the holographic principle, we still need the existence of a locus (the boundary) from which the bulk spacetime will emerge.

On the other hand, if only quantum correlations matter, as suggested in \cite{Silva:2023ieb}, combinatorial information can exist without the necessity of a locus for the existence of objects. In fact, in such a scenario,  localized objects do not exist anyway, but only correlations without objects correlated must correspond to the elements of physical reality.

By taking this route, we must focus on the combinatorial rules governing the structure of quantum correlations from which spacetime must emerge. To explore such rules, we have an interesting fact pointed out by Girelli and Livine  \cite{Girelli:2005ii} that a vector in the representation space of the unitary group corresponds to an intertwiner with $N$ legs dressed by $SU(2)$ representations.

In this way, the $U(N)$ group elements $\hat{h}_{ij}$ in the Eqs. \eq{correlation1} and \eq{correlation2} can be written in terms of  basis vectors in the representation space as

\begin{equation}
\hat{h}_{ij} = e^{M_{ij}} = \mathbb{I} + M_{ij} + \frac{1}{2}M_{ij}M_{ik} +  ...\;\;\;,  \label{spin-net-exp}
\end{equation}

\noindent where $M_{ij}$ corresponds to a $N$-leg intertwiner belonging to the representation space of $U(N)$. Such an intertwiner corresponds to an abstract spin network.

It introduces a very interesting description of the structure of quantum correlations from which spacetime must emerge, where the $U(N)$ quantum states describing such correlations can be written as a superposition of spin network states. In such a scenario, the geometry of the emergent AdS spacetime can be determined by the combinatorial rules governing spin networks.

An important detail here is that abstract spin networks are graphs labeled by $SU(2)$ representations, in the same way we have in full LQG. However, such objects carry only combinatorial information, without any reference to a background geometry or topology \cite{Girelli:2005ii}. Consequently, it is possible to say that, according to the scenario introduced in \cite{Silva:2023ieb}, spacetime must be seen as completely emergent from the combinatorial rules governing abstract spin networks.

By considering such a scenario, the following prescription for the emergence of spacetime can be traced in the context of quantum gravity \cite{Silva:2020bnn, Silva:2023ieb}:

\vspace{1cm}

\noindent \parbox{15.7cm}{\textit{A type IIB string theory living in a classical $AdS$ bulk emerges from a superposition of abstract spin networks describing quantum correlations without correlate, in the limit where the number of correlations becomes large.}}

\vspace{5mm}

We highlight that the pre-geometrical regime from which spacetime ought to emerge must occur far from the limit where the number of branes sourcing the bulk geometry is large, since, in this regime, classical spacetime must emerge, according to the $AdS/CFT$ correspondence.

By observing that the number of branes sourcing gravitationaly the AdS spacetime coincides with the number of external legs of the intertwiners in the approach introduced in \cite{Silva:2023ieb}, we see that this reasoning is consistent with that of \cite{Girelli:2005ii}, which addresses the semiclassical limit of a pre-geometrical $U(N)$ matrix theory formulated in terms of $N$-leg intertwiners. In both cases, the semiclassical regime is reached as $N$, the number of intertwiner legs, becomes large. This large $N$ limit reflects a scenario where the number of quantum correlations also becomes large, matching the prescription discussed above.

In this way, in the theory proposed in \cite{Silva:2023ieb}, spacetime must be seen as emergent from quantum correlations. However, the way quantum correlations organize in order to give rise to spacetime emergence has not been proposed in this scenario. 

In the next section, we shall explore the concept of holographic complexity in the context of the holographic principle, which can provide a possible route to the mechanism of spacetime emergence from quantum correlations: the second law of holographic complexity.

\section{A new window for the emergence of spacetime in quantum gravity: the second law of quantum complexity.} \label{sec4}

In the context of quantum information theory, the concept of quantum complexity has gained increasing importance in recent times, presenting capillarity in several branches of knowledge \cite{Bernstein:1993, Cleve:2000, Vazirani:2002, Watrous:2008any}.

Briefly, complexity expresses the hardness of some task in terms of the amount of computational resources needed to accomplish it. In quantum information theory, any quantum state has an associated complexity, and preparing it using unitary operations is a task in itself. In this context, quantum entanglement has been pointed out as the key resource to computational operations \cite{Steane:1997kb, c.p.willamis-2011}.

Among the several applications of the concept of quantum complexity, it has been pointed out that it must have a fundamental role in holography. As the holographic principle has been considered a guide to quantum gravity, quantum complexity might be a useful way to understand important aspects of the quantum nature of gravitational phenomena.

Considering the AdS/CFT formulation of the holographic principle, it has been suggested that the complexity associated with the quantum entanglement structure of conformal quantum fields on the boundary of an AdS space might give us insight into the emergence of the geometrical properties in the bulk. \cite{Alishahiha:2015rta}.

The origin of such an argument is the fact that entanglement alone is not enough to explain how complex bulk geometries emerge \cite{Susskind:2014moa, Stanford:2014jda}. However, by connecting the bulk spacetime properties to the complexity of quantum states, it is possible to trace an alternative way to understand how larger and more intricate bulk geometries can appear.

Such an idea can be captured by the two main proposals to calculate quantum complexity in the context of the AdS/CFT correspondence, the "\textit{Complexity equals to volume}" \cite{Susskind:2014moa, Stanford:2014jda}, and  the  "\textit{Complexity equals to action}"  \cite{Brown:2015bva} conjectures.

The first proposal suggests that quantum complexity must be related to a volume in the AdS bulk, being proportional to the volume of the maximal codimension-1 bulk slice. The second proposal says that the complexity of a quantum state on the AdS boundary must be related to the gravitational dynamics in the bulk spacetime, being proportional to the action in the Wheeler-DeWitt patch.

Concerning the  "\textit{Complexity equals to volume}" proposal,  the holographic complexity $C_{h}$ related to some region on the AdS boundary is given by \cite{Alishahiha:2015rta}:

\begin{equation}
C_{h} = \frac{Vol(\gamma)}{8\pi L G} = \frac{L^{d-1}}{8(d-1)\pi}\frac{Vol(R)}{\epsilon^{d-1}}, \label{q-complexity}
\end{equation}

\noindent where $L$ is the AdS radius, $\gamma$ is the Ryu-Takayanagi (RT) surface \cite{Ryu:2006bv, Ryu:2006ef}, and $Vol(\gamma)$ is the volume under such a surface in the AdS bulk. The parameter $\epsilon$ denotes a necessary UV cutoff length scale in the boundary theory, and $Vol(R)$ denotes the volume of the boundary spacelike subregion, with a radial size $R$, which goes as $R^{d-1}$. Here $d$ gives the number of bulk spatial dimensions.

Considering the complexity measure above, if quantum complexity increases, bulk geometry also expands. In such a context, an important aspect is that quantum complexity must obey a second law analogous to the thermodynamic one for entropy. According to such a law, if the computational complexity is less than the maximum, then with overwhelming likelihood it will increase till it saturates at a maximal value \cite{Brown:2017jil, Russo:2021hfc, Chapman:2021jbh}.

The second law of quantum complexity must play a crucial role in the issue of the emergence of spacetime since it must dictate the way spacetime must emerge from quantum information. In the present context, it could tell us how spacetime emerges from quantum correlations. In fact, even more implying that spacetime consists of an emergent entity, the second law of quantum complexity implies that it must emerge in a deterministic way based on the rules of quantum computation.

It is also interesting to point out that the second law of quantum complexity must have a crucial role in the investigation of the evolution of the cosmos due to its relationship with the second law of thermodynamics. For example, it has been argued that the cosmic accelerated expansion can be understood in terms of the universe's quantum complexity growth with time \cite{Ge:2017rak}.

In the next sections, we shall propose that some light can be shed on the issue of the Hubble tension if one makes use of the perspective on the emergence of spacetime from quantum correlations introduced in \cite{Silva:2020bnn, Silva:2023ieb}, by considering the second law of holographic complexity as giving the way quantum correlations organize to accomplish such a task.

\section{A relationship between $H_{0}$ and holographic complexity in a quantum gravity scenario.} \label{sec5}

By considering the fundamental aspect of the theory introduced in \cite{Silva:2020bnn, Silva:2023ieb}, that the degrees of freedom to the emergence of spacetime consist of quantum correlations only, the concept of quantum complexity must be helpful.

In such a situation, quantum states and unitary transformations would not be thought of as "things" but as patterns of connectivity and transformation within
the structure of quantum correlations, where the growth of complexity would reflect
structural change, rather than mere dynamical evolution.

Taking into account the issue of the emergence of spacetime, one can consider that quantum entanglement carried by the structure of correlations can be used as the key resource for the quantum computational process driving such a phenomenon \cite{Steane:1997kb, c.p.willamis-2011}. 
In this case, quantum complexity can be given in terms of the number of quantum correlations effectively used to drive spacetime emergence, and the way they organize to accomplish such a task.

To obtain such a number of quantum correlations, one must at first fix both the ultraviolet (UV) and infrared (IR) limits of the theory \cite{Ramallo:2013bua}.
Concerning the IR limit, one must put the system into a region with a finite size, which must be large enough to accommodate the necessary holographic information for the emergence of spacetime.

Particularly, in the $AdS/CFT$ framework, causal diamonds defined on the $AdS$ boundary carry such a necessary amount of information \cite{Hubeny:2012wa, Freivogel:2013zta, Headrick:2014cta}, being considered as the natural unit of holography in a region of spacetime \cite{Krishnan:2019ygy}.
Based on such a fact, we shall take the universe's causal diamond as the region of finite size used to regularize the IR regime of the present theory for the emergence of spacetime (see Fig. \eq{fig1}). 

The radius of the universe's causal diamond is given by

\begin{eqnarray}
R_{\Diamond} &=& \frac{1}{H_{0}}\int^{\infty}_{-1}\frac{dy}{\sqrt{\Omega_{m0}(1+y)^{3} + \Omega_{\Lambda 0}}}   \sim \frac{1}{H_{0}} \;, \label{H-P1}
\end{eqnarray}

\noindent where $\Omega_{m0}$ and  $\Omega_{\Lambda 0}$ correspond to the matter and cosmological constant density parameters, respectively, whose values have been fixed by cosmological observations \cite{Planck:2018vyg}.

On the other hand, we observe that in order to preserve the universe's causal diamond as the unit of holography used for the emergence of the AdS spacetime, we need such a region, defined on the AdS boundary, to be free from extra-dimensional effects. In this sense, we must bound the size of the universe's causal diamond as

\begin{equation}
R_{\Diamond} < L, \label{r-upperbound}
\end{equation}

\noindent since, for $R_{\Diamond} \geq L$, global extra-dimensional effects become important, and we would probe the full extent of the AdS space, leading to distortions in the boundary-to-bulk correspondence. In this way, the condition \eq{r-upperbound} above defines the regime in which the holographic dictionary applies, imposing an upper bound on the holographic information available to the emergence of spacetime.

\vspace{1cm}

\begin{figure}[htb]
\centering  
\includegraphics[width= 10cm, height=8cm]{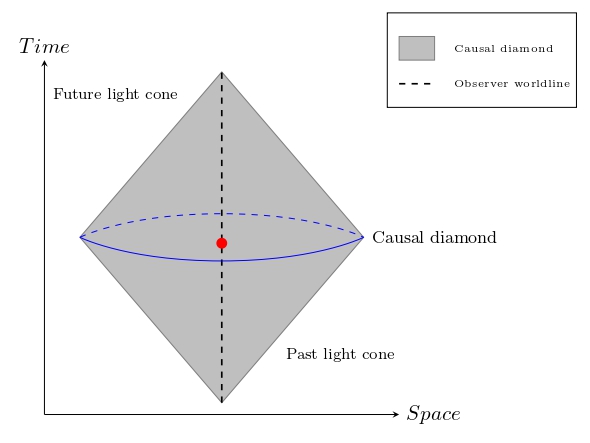}
\caption{Schematic illustration of the universe’s causal diamond: the overlap between the past and future light cones of an observer’s worldline. It shows the region of spacetime that can both influence and be influenced by an observer.} \label{fig1}
\end{figure}

Now, to regularize the UV regime, one must introduce a cutoff length scale in the theory. In the scenario proposed in \cite{Silva:2020bnn, Silva:2023ieb}, the cutoff length scale can be provided by the area parameter $\Delta$ defined by Eq. \eq{delta-g1}, with $\epsilon = \Delta^{1/2}$ as the necessary UV cutoff.


The introduction of such a UV cutoff is possible if one takes into account the usual LQG description of a quantum surface, where each spin network leg patches a quantum of area on the surface. In the present case, such a quantum surface corresponds to the universe's causal diamond horizon.

In this way, let us consider a superposition of spin networks describing the universe's causal diamond, where each spin network state in the superposition is characterized by a specific value of the quantum of area $\Delta$, which defines the measurement scale used to probe the emergent geometry, according to Eq. \eq{delta-g1}.

In this scenario, if one takes $\bar{a}$ as the average quantum of area defined on such a superposition, the number of spin network legs puncturing the causal diamond horizon can be written as

\begin{equation}
N  \sim \frac{A_{\Diamond}}{\bar{a}} ,
\end{equation}

\noindent where $A_{\Diamond}$ is the area of the horizon of the universe's causal diamond.

Now, let us take $\Delta = (1+\beta)\bar{a}$, with $\beta$ measuring the deviation of the quantum of area $\Delta$, characterizing some spin network state, from $\bar{a}$. In other words, $\beta$ can be seen as a quantum fluctuation, or a quantum noise, in the measurement of the area eigenvalues.

\vspace{5mm}

In this case, we can write

\begin{equation}
N  \sim \frac{(1+\beta) A_{\Diamond}}{\Delta} .\label{N-A-beta}
\end{equation}

\vspace{5mm}

\noindent The expression above links the microscopic picture from which spacetime must emerge to the current observational universe by connecting the IR and the UV cutoffs. We must note that a similar result can be obtained by considering the existence of residual higher-dimensional effects in the definition of the universe's causal diamond horizon. It is in agreement with the spirit of the BHHC, where higher-dimensional effects in the bulk are translated as quantum corrections in the four-dimensional gravitational theory induced on the AdS boundary.

By considering the expression \eq{N-A-beta}, for a matter of algebraic convenience, we shall rewrite it as

\begin{equation}
N  \sim   \Big(\frac{A_{\Diamond}}{\Delta}\Big)^{\varepsilon}, \label{N-A}
\end{equation}

\vspace{5mm}

\noindent where the exponent $\varepsilon$ is given by

\vspace{5mm}

\begin{equation}
\varepsilon = 1 + \frac{ln[(1+\beta)]}{ln[N/(1+\beta)]}. \label{epsilon-eq}
\end{equation} 

\vspace{5mm}


\vspace{5mm}

In the present context, the exponent $\varepsilon$ can be interpreted as an effective (fractal) informational dimension of  the surface microstate space, which delimits the amount of quantum information that can be used for the emergence of spacetime, despite the existence of quantum noise. We note that $\varepsilon$ approaches 1 for $\beta \rightarrow 0$, corresponding to the minimal quantum noise in the measurement of the emergent area eigenvalues. Such a limit will be discussed later.


%
%
%
%

By choosing the IR and UV cutoffs above, a superposition of spin networks describing the universe's causal diamond will be considered as the unit of holography in the present context, accommodating all the holographic information necessary for the emergence of spacetime, despite the existence of quantum noise.

In such a scenario, the holographic complexity necessary to accomplish the emergence of cosmic spacetime can be written as

\begin{equation}
C_{h} \sim L^{3}\frac{R_{\Diamond}^{3}}{(\Delta^{1/2})^{3}}, \label{complexity}
\end{equation}

\noindent which corresponds to the holographic complexity \eq{q-complexity} associated with the universe's causal diamond, for the case of $d = 4$, and a UV cutoff length scale $\epsilon = \Delta^{1/2}$.

Now, by considering the relationship between $C_{h}$ and the number of quantum correlations effectively used in the emergence of cosmological spacetime, we shall take:

\begin{equation}
C_{h}  \sim N^{\alpha}. \label{C-A}
\end{equation}

\vspace{5mm}

In the expression above, the exponent $\alpha$ will quantify how efficiently quantum correlations organize to accomplish the emergence of classical spacetime. As $\alpha$ increases, quantum correlations organize more efficiently, which enables the emergence of more complex geometries. On the other hand, as we shall see, there will be an upper bound to such an efficiency, which provides a saturation limit to the complexity associated with the emergent spacetime.

By tracing the basic assumptions of our construction, let us now present our results. From Eqs. \eq{H-P1} and \eq{N-A}, we can find that

\begin{eqnarray}
R_{\Diamond}^{2}H_{0}^{2} &\sim& \textrm{constant}  \nonumber \\
\nonumber \\
N &\sim& \;\; \Big(\frac{R_{\Diamond}^{2}}{\Delta}\Big)^{\varepsilon}.
\end{eqnarray}

\vspace{5mm}

\noindent Such results together provide

\begin{equation}
N^{1/\varepsilon} \Delta H_{0}^{2} \sim \textrm{constant},
\end{equation}

\vspace{1cm}

\noindent which by considering  Eq. \eq{delta-g1} gives

\begin{equation}
N^{\frac{(1 - \varepsilon)}{\varepsilon}} L^{4} H_{0}^{2} \sim \textrm{constant}. \label{eq-1}
\end{equation}

\vspace{1cm}

Now, by taking into account Eqs. \eq{complexity} and \eq{N-A}, we obtain

\begin{equation}
L \sim \frac{C_{h}^{1/3}}{N^{\frac{1}{2\varepsilon}}}. \label{eq-2}
\end{equation}

\vspace{1cm}

By inserting such result in Eq. \eq{eq-1}, one obtains

\begin{equation}
H_{0}^{2} \sim N^{\frac{1+\varepsilon}{\varepsilon}}C_{h}^{-4/3}. \label{H-N-C}
\end{equation}

\vspace{1cm}

Finally, by using Eqs. \eq{complexity}, \eq{H-N-C}, and \eq{C-A}, we obtain

\vspace{5mm}

\begin{equation}
H_{0}^{2} = D C_{h}^{\gamma}  = D N^{\gamma\alpha} = D e^{\alpha \log{N^{\gamma}}} \;, \label{H-C}
\end{equation}

\vspace{5mm}

\noindent where we have englobed all the constants in the calculations above inside the factor $D$.

\vspace{5mm}

In this way, $H_{0}$ depends on the holographic complexity related to the emergence of cosmological spacetime through an exponent $\gamma$ given by

\begin{equation}
\gamma = \frac{3 + 3\varepsilon - 4\alpha\varepsilon}{3\alpha\varepsilon},  \label{gamma-eq}
\end{equation}

\noindent which will correspond to a cosmological scaling to quantum complexity.

\vspace{5mm}


Before proceeding with the analysis of the result above for the exponent $\gamma$, we note first that from Eqs. \eq{C-A} and \eq{eq-2}, we obtain

\begin{equation}
L \sim N^{\frac{2\alpha\varepsilon - 3}{6\varepsilon}}. \label{L-N}
\end{equation}

\vspace{5mm}

\noindent To have consistency with the AdS/CFT correspondence, $L$ must depend on a positive power of $N$ in the sense of obtaining a low curvature regime when $N$ is large \cite{Ramallo:2013bua}.
In this case, we must have $2\alpha\varepsilon - 3 > 0$.


%
%

%
%
%
%
%

Now, the Eqs. \eq{delta-g1}, \eq{N-A} and \eq{L-N},  together give

\begin{equation}
\frac{R_{\Diamond}}{L} \sim N^{\frac{2\alpha -3}{6}}.
\end{equation}

%
%

\noindent By considering such a result, we note that the condition \eq{r-upperbound} imposes that $2\alpha -3 < 0$ in the equation above, and consequently

\vspace{5mm}

\begin{equation}
\alpha < \frac{3}{2}.
\end{equation}

\vspace{5mm}

In this way, the conditions to have the validity of the holographic principle in the present scenario, with the universe's causal diamond as the unity of holography, impose a finite upper bound on the exponent $\alpha$, and then on the efficiency of quantum correlations to organize in order to accomplish the emergence of spacetime.

From the Eq \eq{C-A}, it will also imply a saturation limit for the holographic complexity associated with the emergent spacetime, in the situation where one has a fixed quantum correlation budget $N$. It is in agreement with the second law of quantum complexity as stated above.

In summary, in the present scenario, quantum complexity saturation follows from holographic bounds when one considers the universe's causal diamond as the unity of holography, where the saturation of complexity occurs when the capacity of the causal diamond to accomodate quantum informations reaches its maximum.



By considering the exponent $\gamma$, we have that the only possible solution to it, given by Eq. \eq{gamma-eq}, that is consistent with the existence of an upper bound to the exponent $\alpha$ is

\begin{equation}
\gamma = \frac{3 + 3\varepsilon - 4\alpha\varepsilon}{3\alpha\varepsilon} \geq 0. \label{condition-ii}
\end{equation}

\vspace{5mm}

In fact, the results above provide:

\begin{eqnarray}
\frac{3}{2\varepsilon} < &\alpha& \leq \frac{3(\varepsilon+1)}{4\varepsilon} < \frac{3}{2}, \label{alpha-interval}
\end{eqnarray}

\vspace{5mm}

\noindent with 

\begin{equation}
\varepsilon >  1.
\end{equation}

\vspace{5mm}


%
%
%
%
%
%
\vspace{5mm}

In this way, according to Eqs. \eq{H-C} and \eq{condition-ii}, $H_{0}$ must depend on a non-negative power of the holographic complexity. It corresponds to the central result of our work, which can introduce a new understanding of the very nature of $H_{0}$ based on quantum gravity: that $H_{0}$ must be related to the amount of quantum complexity required for cosmological spacetime comes to the existence.

In this scenario, by considering the second law of holographic complexity, $H_{0}$ must increase during the evolution of the cosmos till it saturates at a maximal value. It must shed some light on the issue of Hubble tension by tracing a way $H_{0}$ can have different values throughout the universe's evolution.

Before proceeding, we note at first that the factor $log(N^{\gamma})$ in Eq. \eq{H-C} corresponds to a logarithmic measure of the effective system size, rescaled by
the cosmological scaling $\gamma$.
To simplify our discussions, we shall fix $N^{\gamma} = e$ in this equation, in a way that

\begin{equation}
H_{0} = De^{\alpha} \label{H-C-1}\;.
\end{equation}

\vspace{5mm}

\noindent It corresponds to fixing the available complexity phase space of the universe, in a way that the behavior of $H_{0}$ will be modeled entirely by the exponent $\alpha$ in the Eq. \eq{H-C-1}. In this case, only the efficiency of complexity growth will control the cosmological expansion. It will be important for the analysis presented in the next sections.


\section{Evolution of holographic complexity.} \label{sec6}

\hspace{7mm} In this section, we shall analyze the evolution of the universe's holographic complexity, by considering a fixed quantum correlation budget, under a given cosmological scale, $N^{\gamma}$. In this way, we shall focus on the evolution of the exponent $\alpha$ in Eq. \eq{C-A}.

In order to link our analysis to the second law of holographic complexity, one can associate the exponent $\alpha$ with a probability weight that indicates how likely it is for high-complexity geometrical states - which are the most likely according to the Second Law - to emerge from quantum correlations.

Now, to apply these ideas to the description of a quantum spacetime given by a superposition of spin network states, each of them characterized by a quantum of area $\Delta$, we note at first that each spin network state is, by its own, a superposition of many quantum geometrical states. Consequently, we shall have not only a simple statistical distribution, but a superposition of many.

In this case, by considering the fluctuation of the parameter $\Delta$ when we change the scale used to define the emergent quantum geometry, in order to relate the exponent $\alpha$ to a probability weight, we must use a superstatistics formulation \cite{Beck:2003kz, Abe:2007}. In this way, we shall write:

\begin{equation}
\alpha(k) = \int_{k}^{\infty} f(k')e^{-k'\Delta}dk'.
\end{equation}

\vspace{5mm}

In the expression above, the steepness parameter $k$  will give the level of coarse-graining of the emergent quantum geometrical states, where large positive values of $k$ strongly suppress larger quantum area eigenvalues, while large negative $k$ values favor them. In the case of $k = 0$, there is no suppression. 
In other words, a larger positive $k$ corresponds to a more sharply-localized, fine-grained behavior. On the other hand, a smaller positive $k$ corresponds to a slower decay: it allows more variance in area contributions, implying a less fine-grained surface structure. Negative $k$ values make coarse-graining effects dominant.

\vspace{5mm}

In this way, by taking $f(k)$ as a Gaussion distribution,

\begin{equation}
f(k) = Ce^{-ak'^{2} - bk'},
\end{equation}

\noindent we obtain

\begin{eqnarray}
\alpha(k) &=& \frac{C}{2} \sqrt{\frac{\pi}{a}} \, e^{\frac{(b + \Delta)^2}{4a}} \, \textrm{erfc}\left( \sqrt{a}\left(k + \frac{b + \Delta}{2a} \right) \right) \nonumber \\
               &=& C' \, \textrm{erfc}\left( \sqrt{a}\left(k + \frac{b + \Delta}{2a} \right) \right), \label{S-alpha}
\end{eqnarray}

\vspace{5mm}

\noindent where $\textrm{erfc}$ is the complementary error function. In the expression above, $a$, $b$, and $C$ are constants. Moreover, we have introduced 

\begin{equation}
C' = \frac{C}{2} \sqrt{\frac{\pi}{a}} \, e^{\frac{(b + \Delta)^2}{4a}},
\end{equation}

\noindent as a constant, for a fixed value of $\Delta$.

\vspace{5mm}

From the result in Eq. \eq{S-alpha}, $\alpha(k)$ will correspond to an S-shaped function (a sigmoid function) of the coarse-graining parameter $k$. The behavior of $\alpha$ is shown in Fig. \eq{fig2}.   

\vspace{5mm}

\begin{figure}[h!]
    \centering
    \begin{subfigure}[t]{1.0\textwidth}
        \centering
        \includegraphics[width=\textwidth]{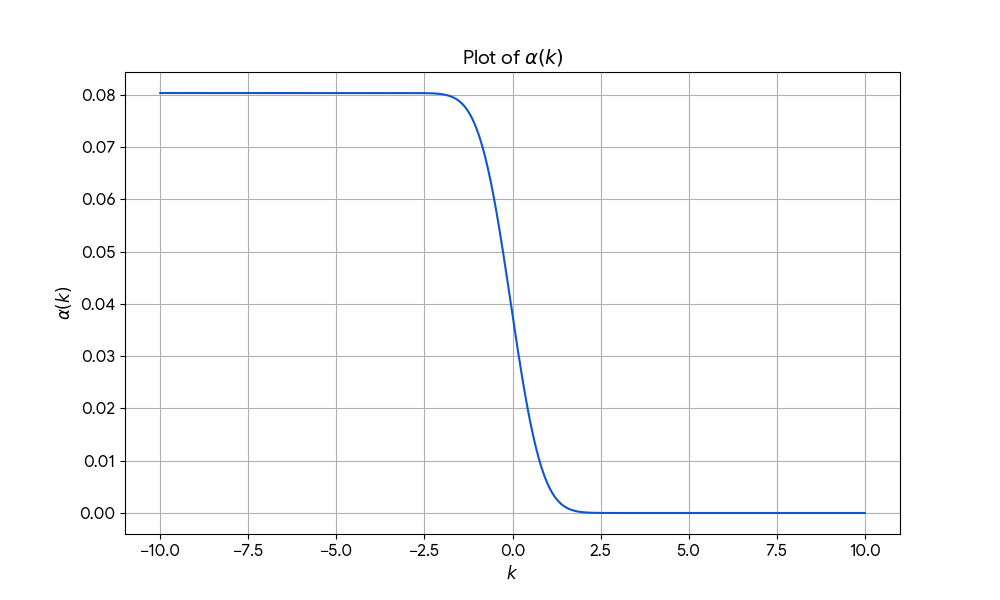}
    \end{subfigure}
    \hfill
    \begin{subfigure}[t]{1\textwidth}
        \centering
        \includegraphics[width=\textwidth]{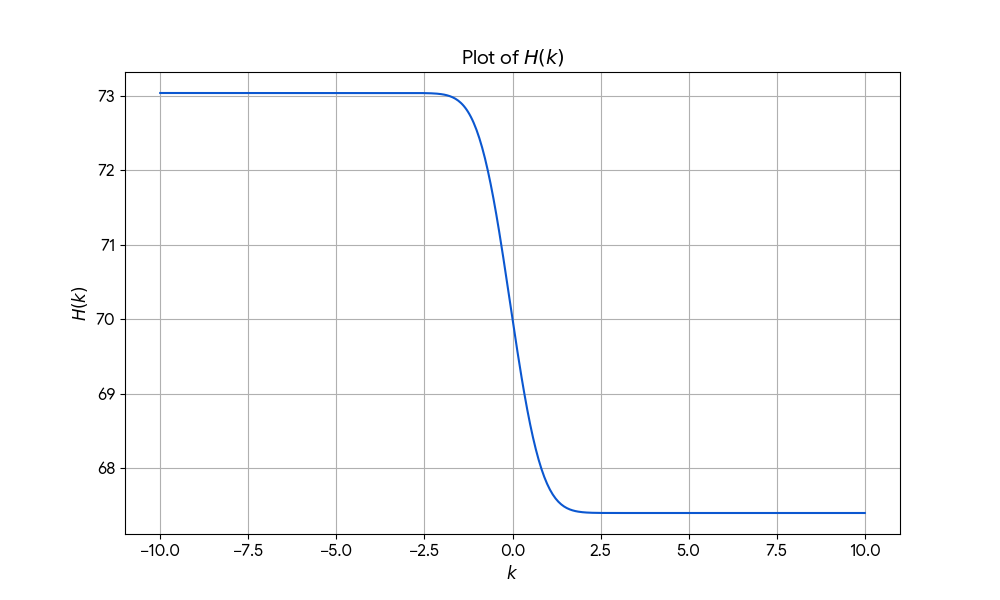}
    \end{subfigure}
    \caption{The exponent $\alpha$, and $H_{0}(k)$ are plotted against the quantum number $k$, by considering $C' = 0.5\; ln (\frac{73.04}{67.4}))$, $D = 67.4$, $a = 1.0$, $b = 0.0$, and $\Delta = 0.1$. Large positive $k$ values strongly suppress larger area eigenvalues corresponding to a more sharply-localized, fine-grained behavior. On the other hand, large $k$ negative values favor larger area eigenvalues, implying a coarse-grained surface structure. }
    \label{fig2}
\end{figure}


As we can observe, $\alpha$ approaches its maximal value when $k$ is negative, and coarse-graining effects become dominant. By considering Eq. \eq{alpha-interval}, it must occur when the exponent $\varepsilon$ approaches 1 from above, corresponding to a reduction of the quantum noise in the measurement of the emergent area eigenvalues. In this case, as coarse-graining effects become dominant, they must filter out the quantum noise associated with the emergence of spacetime, enhancing the efficiency of this process.

Consequently, the reduction of microscopic noise related to spacetime emergence, due to coarse-graining effects, guides the increase of the holographic complexity related to the emergent geometry. In such a scenario, the saturation of holographic complexity must occur when coarse-graining eliminates as much noise as possible, thereby reaching the upper limit of the efficiency of spacetime emergence from quantum correlations. This is in line with some results related to holographic coarse-graining presented in the literature \cite{Behr:2015yna, Behr:2015aat, Murdia:2020iac, Guijosa:2022jdo, Lin:2023jah}.

It brings us a very interesting result in the present context: that larger quantum area eigenvalues are shown to be more efficient at processing quantum information toward the emergence of classical geometry. It is in agreement with some ideas on holography in a quantum spacetime where area size is viewed as a measure of the efficiency in processing holographic quantum information \cite{Markopoulou:1999iq}.

%
%
%
%


\section{Evolution of $H_{0}$ through cosmological history.} \label{sec7}

\hspace{0.7cm} Now, let us consider the consequences of the discussions above to the behavior of the Hubble-Lema\^{i}tre parameter.

As we can observe, from Eq. \eq{H-C-1},  the results obtained for the exponent $\alpha$ also leads to an S-shaped behavior of $H_{0}$ (Fig. \eq{fig2}). It can reproduce the measured values by early and late-time observations, by a suitable choice of the constants. 

In this case, the early-time value occurs at higher redshifts, when the universe was young, and fine-grained emergent quantum geometrical states are dominant, in a way that higher quantum noise effects are more abundant. On the other hand, the late-time values occur at low redshifts when coarse-grained emergent geometrical states become dominant, reducing microscopic noise to a minimum.

Intermediate redshift values of $H_{0}$ can also be fitted in this scenario. In this way, by taking $k = k(z)$, we have that it must change its sign above and below some critical value of the redshift $z_{c}$.
In this case, one can likewise expand

\begin{eqnarray}
k(z) &&= k_{0}(z-z_{c}) + \textrm{fluctuation terms} \nonumber \\   
       &&\approx  \; k_{0}(z-z_{c}), 
\end{eqnarray}

\noindent where it is assumed that $k(z) > 0$ for the high-redshift phase, while $k(z) < 0$ for the low-redshift phase, for a transition to occur. It corresponds to the Landau theory of phase transitions \cite{Landau-1937}.

Based on such considerations, we have plotted our theoretical result for $H_{0}$ in Fig. \eq{fig3}, by assuming $k_{0} = 20$ and $z_{c} = 0.4$.
The plot shows the fitting of our theoretical result with the key measured values for $H_{0}$. In this way, our model can fit the early, late, and middle-time measurements of $H_{0}$, being in agreement with observations.

\begin{figure}[htb]

\centering 

\includegraphics[width= 13cm, height=10cm]{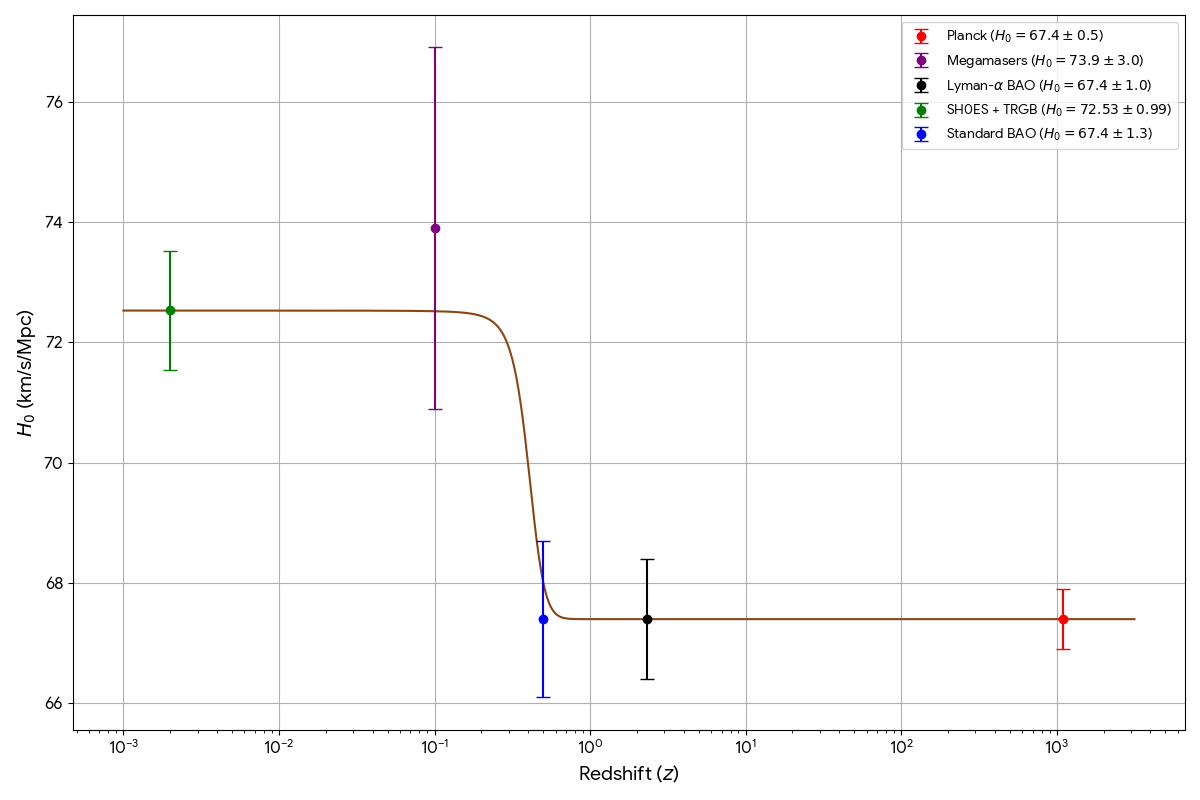}

\caption{The plot shows the $S$-shaped behavior of $H_{0}$ given by our results with $k_{0} = 20$, and $z_{c} = 0.4$. The plot also shows the key early, middle, and late-time measured values for $H_{0}$ against the effective redshift related to the measurement: Planck \cite{Planck:2018vyg}, Standard BAOs \cite{eBOSS:2020yzd}, Ly$\alpha$ forest + BAOs \cite{eBOSS:2020tmo}, SHOES + TRGB \cite{Riess:2021jrx}, and Megamasers \cite{Pesce:2020xfe}. } \label{fig3}

\end{figure}

\section{Conclusions and discussions.} \label{conc}

\hspace{7mm} In the present paper, we utilized the quantum gravity scenario introduced in \cite{Silva:2020bnn, Silva:2023ieb} to approach the important issue of the Hubble tension. In this case, it was possible to trace an interesting relationship between $H_{0}$ and the holographic complexity associated with the emergence of cosmological spacetime.

As an important consequence, we have obtained a link between $H_{0}$ and the second law of quantum complexity.
According to such a law, quantum complexity must increase with time till it saturates at a maximal value \cite{Brown:2017jil, Russo:2021hfc, Chapman:2021jbh}.

In this case, from the results found in the present paper, $H_{0}$ must increase with the universe's holographic complexity during the evolution of the cosmos till it saturates. At this point, such a parameter must reach a constant behavior, recovering the description given by the standard cosmological model. This is in line with several results that have pointed out the possibility that $H_{0}$ must vary throughout the cosmological evolution \cite{Montani:2023xpd, Montani:2023ywn, Schiavone:2022wvq, Dainotti:2021pqg, Dainotti:2022bzg}.

In front of such results, the difference between the values obtained for $H_{0}$ when one considers early and late-time measurements must be only a testimony that our universe has evolved from a relatively simple state at its early times to a very complex place containing structures like galaxies, stars, quasars, black holes, as well as accelerated expansion. In this sense, it is interesting that black holes have been pointed out as the structures that carry the maximal amount of quantum complexity in nature \cite{Barbon:2020olv}.

Such a quantum gravitational reasoning about the Hubble tension can bring us a change of perspective on cosmological observations by taking quantum complexity as a fundamental concept to describe physical reality, with its second law serving as a guide to the emergence of spacetime from quantum correlations.

In this sense, we must note at first that, in the context of the theory introduced in \cite{Silva:2020bnn, Silva:2023ieb}, there is no role for observers (quantum reference frames) in a quantum gravitational context. This is because there is no fundamental distinction between quantum reference frames and quantum correlations in this scenario, where only quantum correlations are led on the table as the fundamental degrees of freedom of the theory.

In this way, our results introduce an observer-independent approach to quantum cosmology, where observer-independence occurs at the fundamental microscopic level, where the cosmos itself is described by quantum correlations structured by the universal law of quantum complexity growth, independent of observers. Consequently, the universe must have a microscopic objective reality based on this law.

On the other hand, observer-dependence can be recovered during the emergence of spacetime, where "observers" must arise as coarse-grained structures, corresponding to different complexity patterns of quantum correlations. In this way, the idea of observer dependence must be replaced by the concept of observer emergence in the scenario proposed in the present paper.


One can catch sight of this by noting that $C_{h}^{-\gamma} \sim H_{0}^{-1}$ gives a distance scale in cosmology, corresponding to the observable universe's radius. Consequently, it is possible to understand the different $H_{0}$ values as a manifestation of the universe's complexity dependence on distance.
In this case, the universe's complexity must decrease when we consider large distance scales.



Therefore, the results found in the present paper can offer not only an understanding of the important issue of the Hubble tension based on quantum gravity but also a new perspective on cosmological observations through a possible link between the micro and the macrostructure of the cosmos: the less holographic complexity we observe, the further into the universe we can see.

In this way, different observers in the universe emerge only as different measures of the holographic complexity associated with the emergence of spacetime by considering different regions of the cosmos.
In such a scenario, when different observers sample the universe, they inherit the same objective quantum complexity law but can translate it into different effective portrayals of the cosmological parameters depending on their complexity budget.
It can open up the range of investigation in quantum cosmology by revealing a new perspective of the world where quantum complexity assumes a central role.

%
%
%
%
%

Future explorations in Astrophysics can be done to explore the scenario introduced in the present work by considering, e.g., gravitational waves \cite{Adhikari:2021ked}, as future detectors become able to measure the quantum complexity amount carried by this kind of signal.

\section*{Acknowledgements}
The author acknowledges the anonymous referees for
 useful comments and suggestions.

\end{document}